\documentclass[aps,pra,twocolumn,showpacs,superscriptaddress,longbibliography]{revtex4-1}
\usepackage{graphicx} 
\usepackage{amsmath}
\usepackage{graphicx,epstopdf}
\usepackage{gensymb}
\newcommand{\be}{\begin{equation}}
\newcommand{\ee}{\end{equation}}
\newcommand{\bea}{\begin{eqnarray}}
\newcommand{\eea}{\end{eqnarray}}
\newcommand{\bse}{\begin{subequations}}
\newcommand{\ese}{\end{subequations}}

\usepackage{color}
\usepackage[colorlinks,bookmarks=false,citecolor=darkblue,linkcolor=red,urlcolor=blue]{hyperref}

\definecolor{darkred}{rgb}{0.7,0.0,0.0}

\definecolor{darkblue}{rgb}{0,0.02,0.45}

\definecolor{darkgreen}{rgb}{0.02,0.45,0.0}

\definecolor{violet}{rgb}{0.8,0.2,0.6}

\begin{document}

\title{Frustration of square cupola in Sr(TiO)Cu$_{4}$(PO$_{4}$)$_{4}$}

\author{S. S. Islam}
\affiliation{School of Physics, Indian Institute of Science
Education and Research Thiruvananthapuram-695016, India}
\author{K. M. Ranjith}
\affiliation{Max Planck Institut fur Chemische Physik fester Stoffe, Nothnitzer Strasse 40, 01187 Dresden, Germany}
\author{M. Baenitz}
\affiliation{Max Planck Institut fur Chemische Physik fester Stoffe, Nothnitzer Strasse 40, 01187 Dresden, Germany}

\author{Y. Skourski}
\affiliation{Dresden High Magnetic Field Laboratory, Helmholtz-Zentrum Dresden-Rossendorf, 01314 Dresden, Germany}

\author{A. A. Tsirlin}
\email{altsirlin@gmail.com}
\affiliation{Experimental Physics VI, Center for Electronic Correlations and Magnetism, University of Augsburg, 86135 Augsburg, Germany}

\author{R. Nath}
\email{rnath@iisertvm.ac.in}
\affiliation{School of Physics, Indian Institute of Science Education and Research Thiruvananthapuram-695016, India}
\date{\today}

\begin{abstract}
The structural and magnetic properties of the square-cupola antiferromagnet Sr(TiO)Cu$_{4}$(PO$_{4}$)$_{4}$ are investigated via x-ray diffraction, magnetization, heat capacity, and $^{31}$P nuclear magnetic resonance experiments on polycrystalline samples, as well as density-functional band-structure calculations. The temperature-dependent unit cell volume could be described well using the Debye approximation with the Debye temperature of $\theta_{\rm D} \simeq $ 550~K. Magnetic response reveals a pronounced two-dimensionality with a magnetic long-range-order below $T_{\rm N} \simeq 6.2$~K. High-field magnetization exhibits a kink at $1/3$ of the saturation magnetization. Asymmetric $^{31}$P NMR spectra clearly suggest strong in-plane anisotropy in the magnetic susceptibility, as anticipated from the crystal structure. From the $^{31}$P NMR shift vs bulk susceptibility plot, the isotropic and axial parts of the hyperfine coupling between $^{31}$P nuclei and the Cu$^{2+}$ spins are calculated to be $A_{\rm hf}^{\rm iso} \simeq 6539$ and $A_{\rm hf}^{\rm ax} \simeq 952$~Oe/$\mu_{\rm B}$, respectively. The low-temperature and low-field $^{31}$P NMR spectra indicate a commensurate antiferromagnetic ordering. Frustrated nature of the compound is inferred from the temperature-dependent $^{31}$P NMR spin-lattice relaxation rate and confirmed by our microscopic analysis that reveals strong frustration of the square cupola by next-nearest-neighbor exchange couplings.
\end{abstract}
\maketitle

\section{Introduction}
\begin{figure*}
	\includegraphics[scale=0.8]{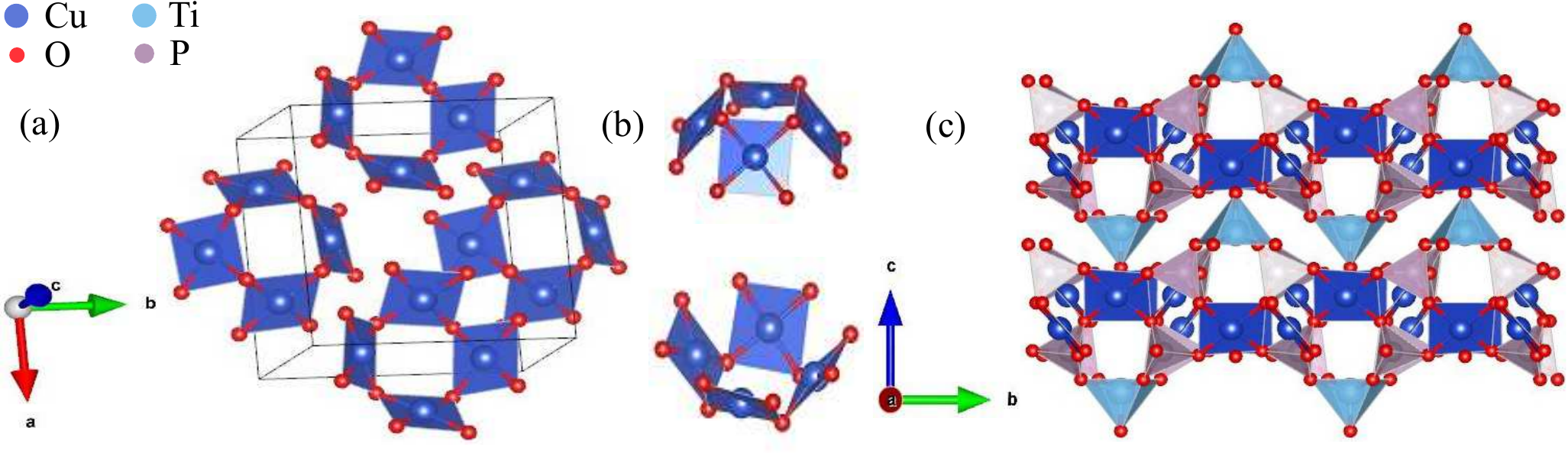}
	\caption{\label{Fig1}(a) Crystal structure of Sr(TiO)Cu$_{4}$(PO$_{4}$)$_{4}$ viewed along with the three principal axes in the left. The Cu and O ions are shown by the blue and red spheres, respectively. The CuO$_{4}$ planes (plaquettes) within the up/down square cupola are shown in blue. The black outline represents the unit cell. (b) The upward and downward Cu$_{4}$O$_{12}$ square cupola are shown. (c) Crystal structure of Sr(TiO)Cu$_{4}$(PO$_{4}$)$_{4}$ along the $a$-axis. The Ti and P atoms are shown by cyan and grey spheres, respectively.}
\end{figure*}
Recently, two new chiral compounds $A$(TiO)Cu$_{4}$(PO$_{4}$)$_{4}$ ($A$ = Ba and Sr) (abbreviated as BTCPO and STCPO, respectively) were synthesized~\cite{Kimura2016IC}. Both the compounds crystallize in the tetragonal space group $P42_{1}2$ with the lattice parameters [$a=9.6028(5)$~\AA\ and $b=7.1209(5)$~\AA] and [$a=9.5182(5)$~\AA\ and $b=7.0087(5)$~\AA], respectively. BTCPO and STCPO feature a quasi-two-dimensional (2D) structure, which is formed by the upward and downward Cu$_{4}$O$_{12}$ cluster-assembly units with alternating orientations in the $ab$-plane, as depicted in Fig.~\ref{Fig1}(a). Each Cu$_{4}$O$_{12}$ unit consists of four corner-sharing CuO$_{4}$ square plaquettes forming a non-coplanar structure called (irregular) $square-cupola$. Two types of the square-cupola units (upward and downward) are present in the structure [see Fig.~\ref{Fig1}(b)]. Each upward Cu$_{4}$O$_{12}$ square cupola is connected to the nearest downward square cupola unit and with the nearest upward TiO$_{5}$ square pyramid [located on top (or bottom) of each upward (or downward) Cu$_{4}$O$_{12}$ square-cupola unit] by the PO$_{4}$ tetrahedra.
 
A detailed magnetoelectric study has been reported for BTCPO via magnetization, heat capacity, dielectric, neutron scattering experiments, and subsequent model calculations~\cite{Kimura2016NComm,Kato2017PRL,Kimura2017PhysicaB,Babkevich2017arXiv}. A huge anomaly in the temperature-dependent dielectric constant was observed at the Ne$\acute{e}$l temperature $T_{\rm N} \simeq 9.5$~K in the magnetic field applied along the [100] and [1$\bar{1}$0] directions. It has been proposed that the asymmetric square-cupola units in the crystal structure can sustain magnetoelectric-active multipole moments associated with an antiferroic magnetic quadrupole order. Later, a combined experimental [high-field magnetization measurement] and theoretical [cluster mean-field approximation] study~\cite{Kato2017PRL} interpreted the magnetoelectric phase diagram and analyzed microscopically how the asymmetric square-cupola units trigger the magnetoelectric response through anisotropic exchange interactions. The critical role of the in-plane component of the Dzyaloshinsky-Moriya (DM) interaction, induced by the non-coplanar structure of the square cupola was established. The non-collinear magnetic structure of BTCPO has been determined using neutron diffraction and, eventually, spherical neutron polarimetry~\cite{Babkevich2017arXiv}. 

Interestingly, the degree of chirality can be tuned in the $A$($B$O)Cu$_{4}$(PO$_{4}$)$_{4}$ ($B=$ Ti and V) series of compounds by changing both the $A$ and $B$ cations. For example, the structural chirality is more pronounced in STCPO compared BTCPO~\cite{Kimura2016IC}. We therefore investigated and report below the magnetic behavior of STCPO, as revealed by different experimental techniques and \textit{ab initio} calculations.

\section{Methods}
Polycrystalline sample of STCPO was synthesized by the conventional solid-state reaction technique. The synthesis involves two steps. In the first step, the initial reactants SrCO$_{3}$ (Aldrich, 99.995\%), CuO (Aldrich, 99.999\%), TiO$_{2}$ (Aldrich, 99.9\%), and (NH$_{4}$)H$_{2}$PO$_{4}$ (Aldrich, 99\%) were taken in stoichiometric ratios, ground thoroughly, and pressed into pellets. The pellets were kept in a quartz crucible and fired at 300~$\degree$C for 12 hours in air to remove NH$_{3}$ and CO$_{2}$ from the sample. In the second step, the sample was re-ground, pelletized, and annealed at 950~$\degree$C for 24 hours in air in a platinum crucible. Phase purity of the sample was confirmed by powder x-ray diffraction (XRD) measurements using the diffractometer from PANalytical (Cu\textit{K}$_{\alpha}$ radiation, $\lambda_{\rm avg}\simeq 1.5418$~{\AA}). The temperature-dependent powder XRD experiments were carried out over a broad temperature range (15~K$\leq T \leq 650$~K) using the Oxford Phenix (for low temperatures) and Anton-Paar HTK 1200N (for high temperatures) attachments.

Magnetization ($M$) measurements were performed using the vibrating sample magnetometer (VSM) attachment to the Physical Property Measurement System (PPMS, Quantum Design). Heat capacity ($C_{\rm p}$) was measured using the heat capacity option in the PPMS, adopting the relaxation technique. High-field magnetization measurement up to 60~T was performed at the Dresden High Magnetic Field Laboratory using pulsed fields. Details of the measurement procedure are described in Ref.~\cite{Tsirlin2009PRB}.

The NMR measurements were carried out using pulsed NMR techniques on $^{31}$P (nuclear spin $I=1/2$ and gyromagnetic ratio $\gamma_{N}/2\pi = 17.237$\,MHz/T) nuclei. We have carried out the measurements at two different radio frequencies of $70.41$\,MHz and $25$\,MHz, which correspond to the applied fields of about $4.085$\,T and $1.45$\,T, respectively. The spectra were obtained by sweeping the field at a fixed frequency. The NMR shift $K(T)=(H_{\rm ref}-H(T))/H(T)$ was determined by measuring the resonance field of the sample [$H(T)$] with respect to the nonmagnetic reference H$_{3}$PO$_{4}$ (resonance field $H_{\rm ref}$). The $^{31}$P spin-lattice relaxation rate $1/T_{1}$ was measured by the conventional inversion recovery method.

Exchange couplings $J_{ij}$ of the spin Hamiltonian
\begin{equation}
 H=\sum_{\langle ij\rangle}J_{ij}\mathbf S_i\mathbf S_j,
\end{equation}
where the summation is over all pairs $\langle ij\rangle$, were obtained from density-functional (DFT) band-structure calculations performed in the full-potential local-orbital \texttt{FPLO} code~\cite{fplo} using Perdew-Burke-Ernzerhof flavor of the exchange-correlation potential~\cite{pbe}. First Brillouin zone was sampled by a $k$-mesh with up to 216 points in the symmetry-irreducible part. Exchange couplings were extracted using two complementary approaches. On one hand, the uncorrelated band structure was parametrized by a tight-binding model, and exchange couplings were estimated via the hopping parameters. On the other hand, total energies from DFT+$U$ calculations were used in a mapping procedure~\cite{xiang2011}. In DFT+$U$, electronic correlations in the Cu $3d$ shell are accounted for on the mean-field level using the on-site Coulomb repulsion $U_d=9.5$\,eV and Hund's coupling $J_d=1$\,eV~\cite{janson2012,nath2013,lebernegg2013}. 

For simulating magnetic susceptibility and magnetization, finite lattices with up to 4 square cupolas and periodic boundary conditions were used. The simulations were performed using the \texttt{fulldiag} routine of the \texttt{ALPS} package~\cite{alps}.

\section{Results and Discussion}
\subsection{Crystal structure}
\begin{figure}
	\includegraphics[scale=1.1] {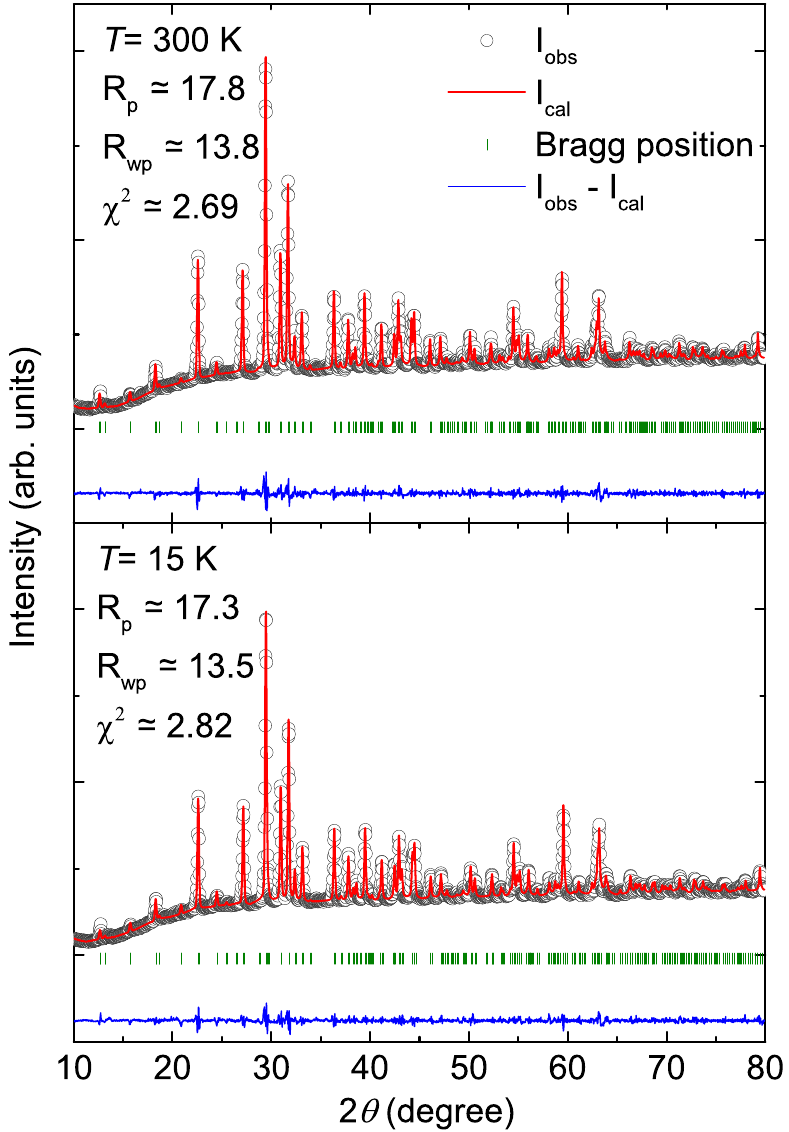}
	\caption{\label{Fig2} X-ray diffraction data collected at $T=300$~K and 15~K. The solid lines denote the Le-Bail fit of the data. The Bragg peak positions are indicated by green vertical bars, the bottom blue line indicates the difference between the experimental and calculated intensities.}
\end{figure}
\begin{figure}
	\includegraphics[scale=1.1] {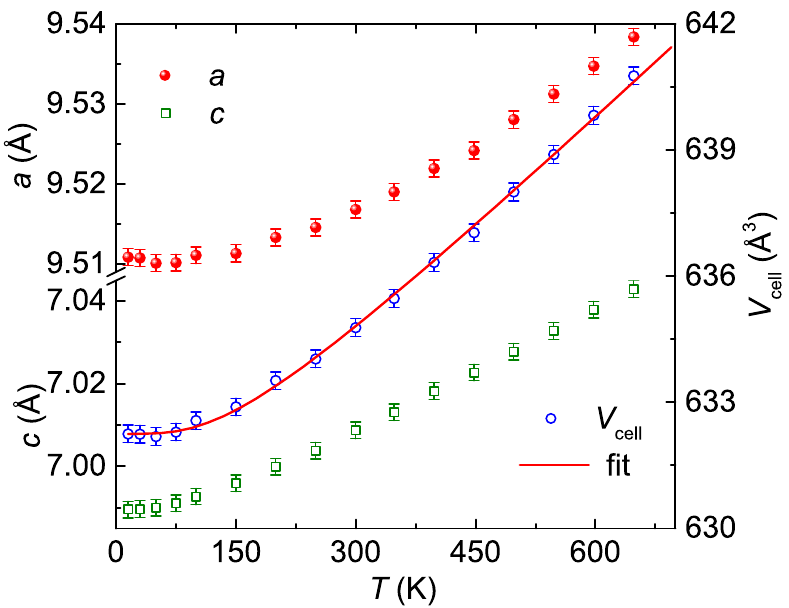}
	\caption{\label{Fig3} Variation of lattice constants ($a$ and $c$) and unit cell volume ($V_{\rm cell}$) with temperature. The solid line represents the fit of $V_{\rm cell}(T)$ by Eq.~\eqref{VcellvsT}.}
\end{figure}
In order to check the phase purity and to detect structural transitions, if any, powder XRD was measured at different temperatures. Le-Bail fits of the XRD patterns were carried out using the \texttt{FullProf} package.\cite{Rodriguez1993PhysicaB} The initial structural parameters for this purpose were taken from Ref.~\cite{Kimura2016IC}. Figure~\ref{Fig2} shows the powder XRD patterns at 300~K and 15~K along with the fits. The absence of unindexed peaks confirms phase purity of our sample.
The obtained lattice constants and volume of the unit cell ($V_{\rm cell}$) at room temperature are $ a = 9.5168(1) $~\text{\AA}, $ c = 7.0087(1) $~\text{\AA}, and $V_{\rm cell} \simeq 634.78(2)$~\text{\AA}$^{3}$, which are consistent with the previous report~\cite{Kimura2016IC}. Figure~\ref{Fig3} displays the temperature variation of lattice constants and unit cell volume. No structural transition was observed down to 15~K. The lattice constants and $V_{\rm cell}$ were found to decrease systematically upon cooling. The temperature variation of $V_{\rm cell}$ was fitted by the equation~\cite{Pakhira2016PRB}
\begin{equation}
V(T)=\gamma U(T)/K_0+V_0,
\label{VcellvsT}
\end{equation}
where $V_0$ is the cell volume at $T = 0$~K, $K_0$ is the bulk modulus, and $\gamma$ is the Gr$\ddot{\rm u}$neisen parameter. $U(T)$ is the internal energy which can be expressed in terms of the Debye approximation as
\begin{equation}
U(T)=9p\,k_{\rm B}T\left(\frac{T}{\theta_{\rm D}}\right)^3\int_{0}^{\theta_{\rm D}/T}\dfrac{x^3}{e^x-1}dx.
\end{equation}
Here, $p$ is the number of atoms in the unit cell, and $k_{\rm B}$ is the Boltzmann constant. Using this approximation (see Fig.~\ref{Fig3}), the Debye temperature ($\theta_{\rm D}$) and other parameters were estimated to be $\theta_{\rm D} \simeq $ 550~K, $\gamma/K_0 \simeq 2.68 \times 10^{-5}$~Pa$^{-1}$, and $V_0 \simeq 632.2$~\AA$^{3}$.

\subsection{Magnetization}
\begin{figure}
	\includegraphics[scale=1.0] {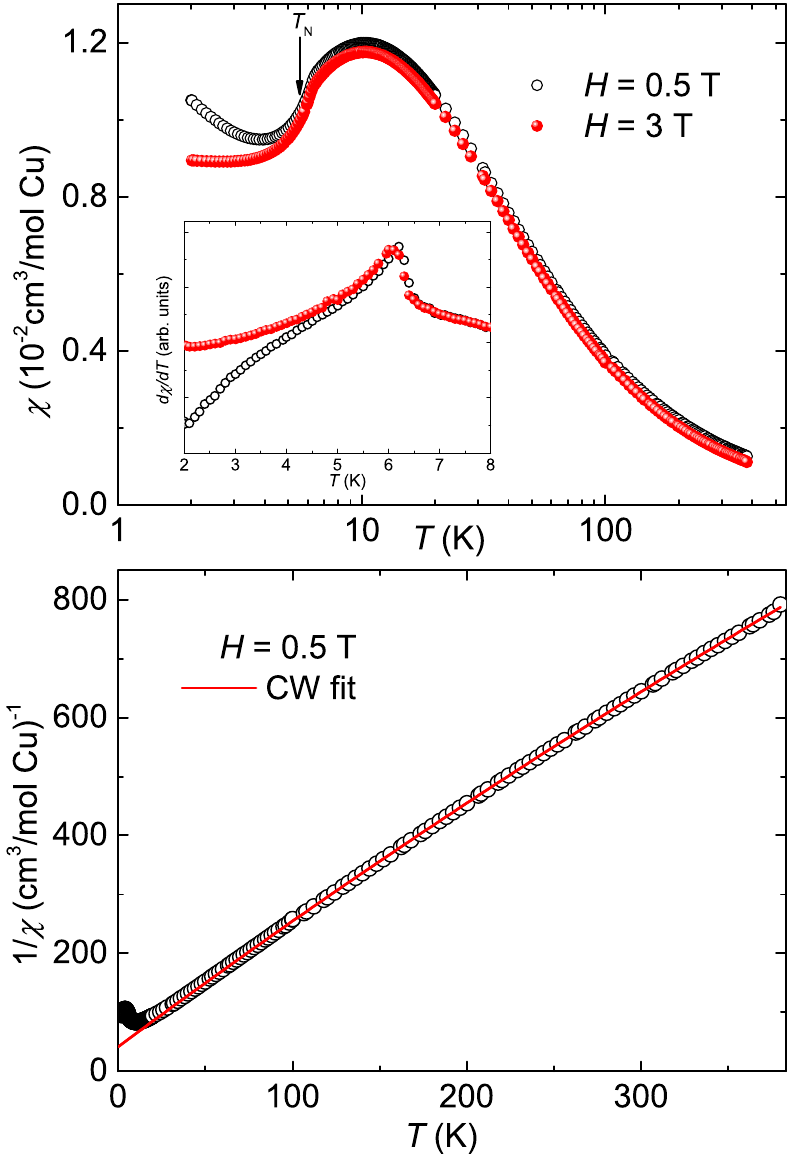}
	\caption{\label{Fig4} Upper panel: temperature-dependent magnetic susceptibility $\chi(T)$ measured at different applied fields. Inset: $d\chi/dT$ vs temperature in the low-temperature regime to highlight the magnetic transition. Lower panel: inverse susceptibility ($1/\chi$) vs temperature and the solid line represents the fit using Eq.~(\ref{cw})}
\end{figure}
The static magnetic susceptibility $\chi(T)$ ($\equiv M/H$) of the polycrystalline STCPO sample measured as a function of temperature at different applied fields $H= 0.5$~T and 3~T is shown in the upper panel of Fig.~\ref{Fig4}. At high temperatures, $\chi(T)$ increases with decreasing temperature in a Curie-Weiss manner and passes through a broad maximum around $T_{\chi}^{\rm max} \simeq 10$~K. This broad maximum is indicative of a short-range order, anticipated for low-dimensional antiferromagnetic (AFM) spin systems. The temperature corresponding to the broad maximum is a measure of the dominant AFM exchange coupling. Below $T_{\chi}^{\rm max}$, $\chi(T)$ exhibits a small kink at $T_{\rm N} \simeq 6.2$~K, a possible indication of magnetic long-range ordering (LRO). In order to see this feature, we plotted temperature derivative of $\chi$ in the inset of Fig.~\ref{Fig4}. It shows a clear anomaly at $T_{\rm N} \simeq 6.2$~K. At very low temperatures, $\chi(T)$ shows an upturn, which is suppressed with the applied magnetic field. This upturn could be due to a small amount of extrinsic paramagnetic impurities and/or defects present in the sample.
 
For a quantitative analysis, the high-temperature $\chi(T)$ data were fitted by the modified Curie-Weiss (CW) law
 \begin{equation}
 \chi(T)=\chi_0+\frac{C}{T+\theta_{\rm CW}}.
 \label{cw}
 \end{equation}
Here, $\chi_0$ is the temperature-independent susceptibility, which includes Van-Vleck paramaganetism and core diamagnetism, and the second term is the CW law. Our fit in the high-temperature regime ($T \geq 100$~K) (see the lower panel of Fig.~\ref{Fig4}) yields the following parameters: $\chi_{0} \simeq 1.42 \times 10^{-4}$~cm$^{3}$/mol-Cu$^{2+}$, the Curie constant $C \simeq 0.45$~cm$^{3}$K/mol-Cu$^{2+}$, and the Curie-Weiss temperature $\theta_{\rm CW} \simeq 18.7$~K. The core diamagnetic susceptibility $\chi_{\rm core}$ of STCPO was calculated to be $-2.72 \times 10^{-4}$~cm$^{3}$/mol by adding the core diamagnetic susceptibilities~\cite{Selwood1956Book} of individual ions: Sr$^{2+}$, Ti$^{4+}$, Cu$^{2+}$, P$^{5+}$, and O$^{2-}$. The Van-Vleck paramagnetic susceptibility $\chi_{\rm VV}$ was estimated by subtracting $\chi_{\rm dia}$ from $\chi_{0}$ to be $\sim 4.14 \times 10^{-4}$~cm$^{3}$/mol. This value of $\chi_{\rm VV}$ is close to the values reported for other cuprates~\cite{Nath2005PRB,Motoyama1996PRL}. 

From the value of $C$, the effective moment is calculated to be $\mu_{\rm eff} \simeq 1.89 \mu_{\rm B}$ using the relation $\mu_{\rm eff} = \sqrt{3k_{\rm B}C/N_{\rm A}}$, where $k_{\rm B}$ is the Boltzmann constant and $N_{\rm A}$ is the Avogadro's number. This value of the effective moment corresponds to a Land$\acute{e}$ $g$-factor of $g\simeq 2.18$ [using $\mu_{\rm eff} = g\sqrt{S(S+1)}\mu_{\rm B}$]. Such a large value of $g$ is typically observed for powder samples containing magnetic Cu$^{2+}$ ions. The positive value of $\theta_{\rm CW}$ suggests that the dominant exchange interactions between Cu$^{2+}$ ions are AFM in nature.

\begin{figure}
	\includegraphics[scale=1.1] {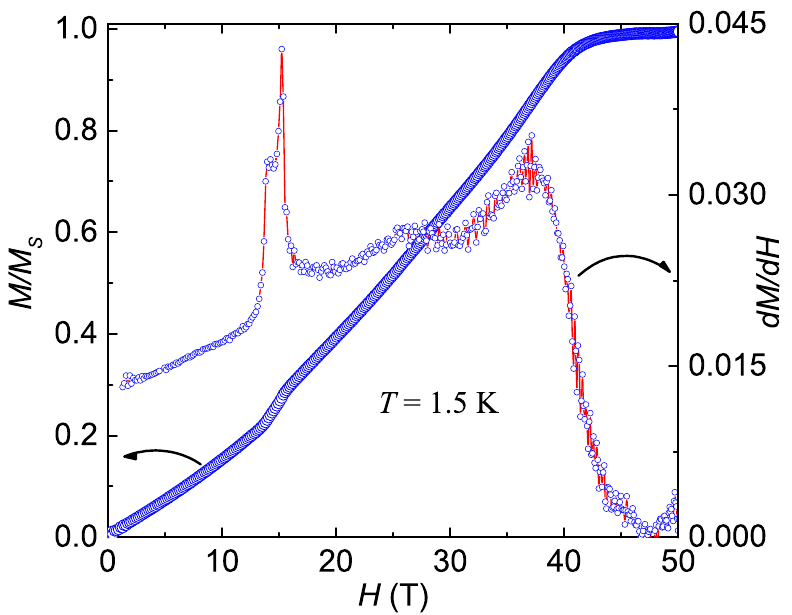}
	\caption{\label{Fig5} Normalized magnetization isotherm ($M$ vs $H$) at $T = 1.5$~K. $dM/dH$ vs $H$ is shown in the right $y$-axis.}
\end{figure}
In order to study field-induced effects, magnetization isotherm ($M$ vs $H$) was measured at $T = 1.5$~K up to 60~T. As shown in Fig.~\ref{Fig5}, $M$ increases almost linearly with $H$ and then exhibits a kink at $H\simeq 15$~T before reaching full saturation at $H_{\rm s} \simeq 40$~T. The kink seems to occur around $\frac13$ of the saturation magnetization ($M_{\rm s}$). To visualize this feature, we plotted the derivative of $M$ with respect to $H$ i.e. $dM/dH$ vs $H$ in the right $y$-axis of Fig.~\ref{Fig5}, which shows a sharp peak at this field. The overall behavior of the $M$ vs $H$ curve is similar to that reported for BTCPO earlier~\cite{Kato2017PRL}.

\subsection{Heat Capacity}
\begin{figure}
	\includegraphics[scale=1] {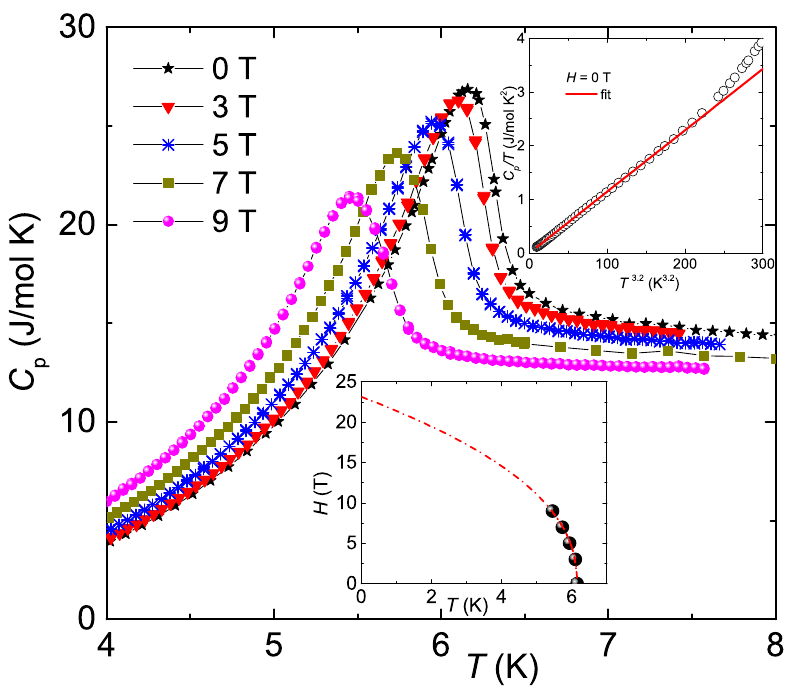}
	\caption{\label{Fig6} $C_p(T)$ of STCPO measured at different applied fields in the low-temperature region around $T_{\rm N}$. Upper inset: $C_p(T)/T$ vs $T^{3.2}$ at zero field, showing the linear regime below $T_{\rm N}$. The solid line is the linear fit. Lower inset: $H$ vs $T_{\rm N}$ phase diagram, with the dash-dotted line showing the fit using the power-law $H(T)=H_{\rm c}\left(1-\frac{T}{T_{\rm N}}\right)^{\beta}$ and the extrapolation of this fit to low temperatures.}
\end{figure}
The temperature-dependent heat capacity $C_{\rm p}(T)$ is presented in Fig.~\ref{Fig6}. At high temperatures, $C_{\rm p}$ is dominated by phonon excitations. As the temperature is lowered, $C_p$ shows a sharp $\lambda$-type anomaly at $T_{\rm N} \simeq 6.2$~K, reflecting the magnetic LRO. Since a non-magnetic analogue compound could not be synthesized, we were unable to extract the magnetic part of the heat capacity from the total $C_{\rm p}(T)$. Hence, no further analysis was possible. As shown in the upper inset of Fig.~\ref{Fig6}, a linear behavior was observed below $T_{\rm N}$ when $C_{\rm p}/T$ is plotted against $T^{3.2}$. Usually, in the three-dimensional (3D) ordered state, the spin-wave dispersion gives rise to a $T^3$ behavior for $C_{\rm p}(T)$. The higher power-law exponent of 4.2 may indicate peculiarities of the magnetic excitation spectrum, such as the presence of a spin gap. However, we were unable to achieve a good fit of $C_{\rm p}(T)$ assuming an activated behavior.

When the magnetic field is applied, the position of the anomaly at $T_{\rm N}$ shifts towards low temperatures, suggesting that the transition is AFM in nature. The variation of $T_{\rm N}$ with $H$ is shown in the lower inset of Fig.~\ref{Fig6}, which is a typical phase diagram expected for an antiferromagnet. Field evolution can be described by an empirical power-law fit $H(T)=H_{\rm c}\left(1-\frac{T}{T_{\rm N}}\right)^{\beta}$ with $H_{\rm c}\simeq 23.1$~T, $T_{\rm N} \simeq 6.15$~K, and $\beta \simeq 0.44$. 

\subsection{$^{31}$P NMR}
The crystal structure of STCPO features one P site. Two adjacent cupolas (one upward and one downward) in the $ab$-plane are connected through the PO$_4$ tetrahedra. Since P is strongly coupled with the Cu$^{2+}$ ions, one can probe static and dynamic properties of Cu$^{2+}$ spins via $^{31}$P NMR.

\subsubsection{$^{31}$P NMR Shift}
\begin{figure}
	\includegraphics [scale=1]{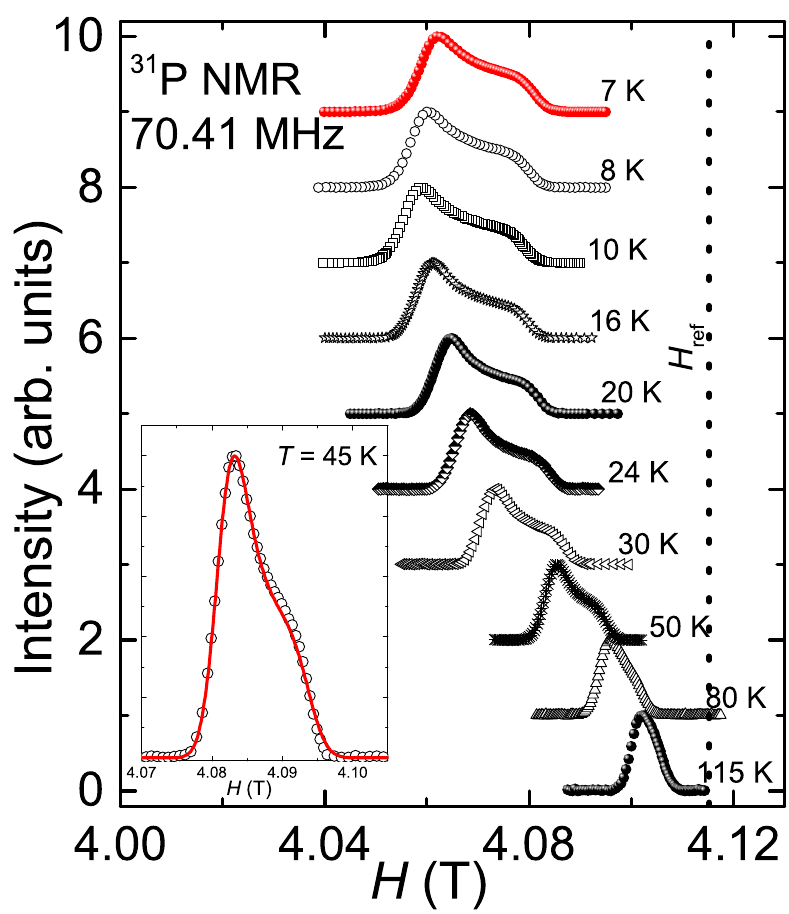}
	\caption{\label{Fig7} (Color online) Field-sweep $^{31}$P NMR spectra at different temperatures $T$ ($T > T_{\rm N}$) for the polycrystalline STCPO sample measured at 70.41\,MHz. The vertical dashed line corresponds to the $^{31}$P resonance frequency of the reference sample H$_{3}$PO$_{4}$. Inset shows the $^{31}$P NMR spectrum at $45$\,K (open circles). The solid line is the fit. The downward arrows point to the anisotropic shift values.}
\end{figure}

\begin{figure}
	\includegraphics [scale=1]{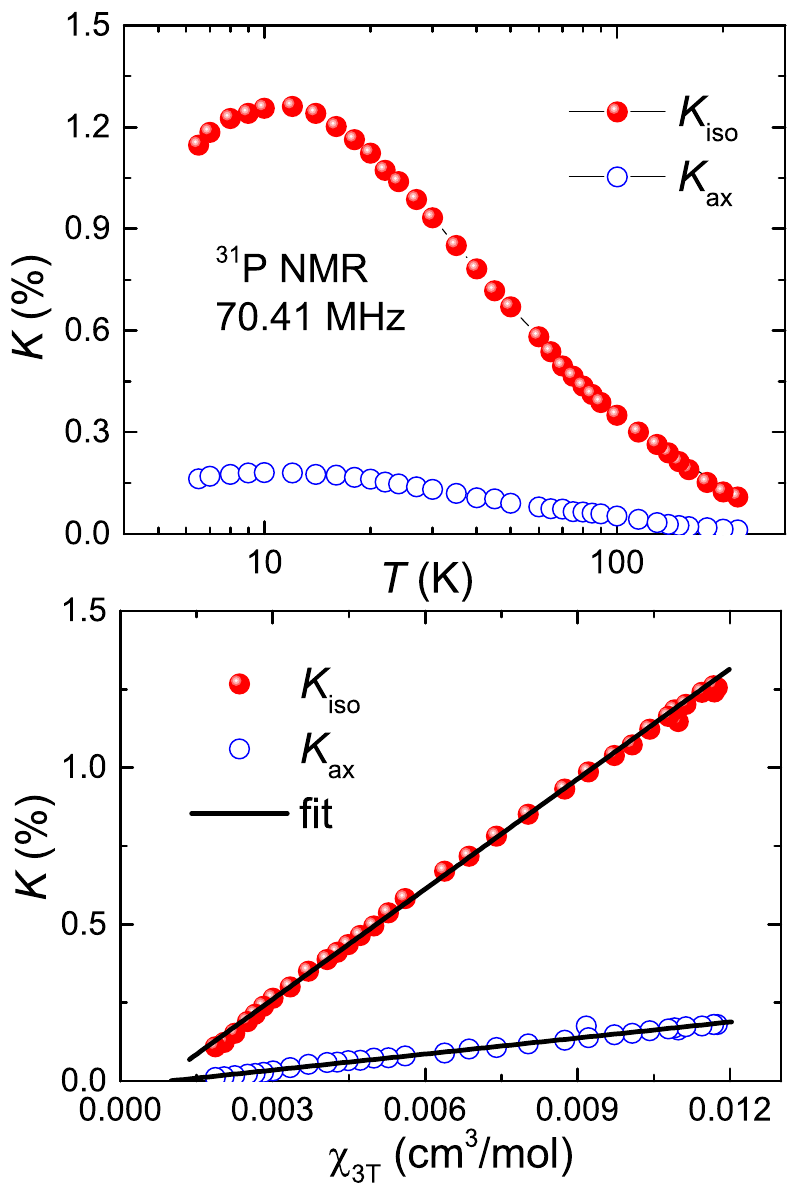}
	\caption{\label{Fig8} (Color online) Upper panel: temperature-dependent isotropic and axial $^{31}$P NMR shifts $K_{\rm iso}$ and $K_{\rm ax}$ vs $T$. Lower panel: $^{31}$P NMR shift vs $\chi$ measured at 3\,T is plotted with temperature as an implicit parameter for both $K_{\rm iso}$ and $K_{\rm ax}$. The solid lines are linear fits.}
\end{figure}

We observed a narrow spectral line above $T_{\rm N}$ as expected for an $I=1/2$ nucleus~\cite{Nath2005PRB,Nath2008PRB}. Figure~\ref{Fig7} presents the $^{31}$P NMR spectra measured at different temperatures. The line shape was found to be asymmetric, similar to that observed for Zn$_2$VO(PO$_4$)$_2$ because of the anisotropy in $\chi(T)$ and/or in the hyperfine coupling constant between the P nucleus and Cu$^{2+}$ spins~\cite{Yogi2015PRB}. The line position is found to shift with temperature. Temperature dependence of the NMR shift $K$ extracted by fitting the spectra (see inset of Fig.~\ref{Fig7}) is presented in Fig.~\ref{Fig8}, which shows a strong anisotropy along different directions. At high temperatures, both isotropic ($K_{\rm iso}$) and axial ($K_{\rm ax}$) parts of the NMR shift vary in a Curie-Weiss manner and then pass through a broad maximum at around 10\,K reflecting the low-dimensional short-range order, similar to the $\chi(T)$ data (Fig.~\ref{Fig4}).

The NMR shift $K(T)$ is a direct measure of the spin susceptibility $\chi _{\rm spin}$ and is free from extrinsic contributions. Therefore, one can write $K(T)$ in terms of $\chi_{\rm spin}(T)$ as
\begin{equation}
	K(T)=K_{0}+\frac{A_{\rm hf}}{N_{\rm A}} \chi_{\rm spin}(T),
	\label{shift}
\end{equation}
where $K_{0}$ is the temperature-independent chemical shift and $A_{\rm hf}$ is the hyperfine coupling constant between the P nuclei and Cu$^{2+}$ electronic spins. The $K$ vs $\chi$ plot with $T$ as an implicit parameter is fitted very well by a straight line [Fig.~\ref{Fig8} (lower panel)] over the whole temperature range ($T > T_{\rm N}$) yielding the isotropic and axial parts of the hyperfine coupling $A_{\rm hf}^{\rm iso} \simeq 6539$ and $A_{\rm hf}^{\rm ax} \simeq 952$~Oe/$\mu_{\rm B}$, respectively.

\begin{figure}
	\includegraphics [scale=1]{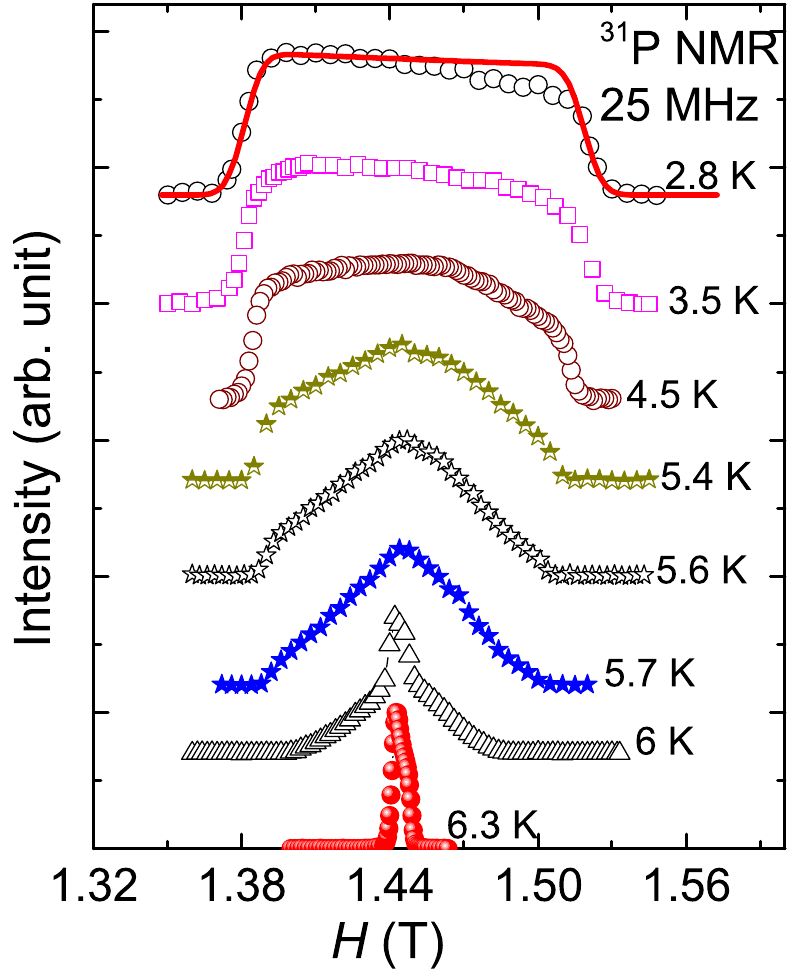}
	\caption{\label{Fig9} (Color online) Temperature-dependent $^{31}$P NMR spectra measured at 25\,MHz below $T_{\rm N}$. The solid line is the fit to the spectrum at $T = 2.8$~K as in Refs.~\onlinecite{Nath2014PRB,Ranjith2016PRB}. The spectra in the paramagnetic state broaden below $T_{N}$ and take a rectangular shape due to the internal field $H_{\rm int}$.}
\end{figure}
\subsubsection{NMR spectra below $T_{\rm N}$}
Below $T_{\rm N}$, the $^{31}$P spectra measured at 70.41\,MHz were found to broaden abruptly and take a rectangular shape at very low temperatures. In order to precisely probe the intrinsic line shape, we re-measured the $^{31}$P spectra at a lower frequency of 25\,MHz. As shown in Fig.~\ref{Fig9}, the $^{31}$P line above $T_{\rm N}$ remains narrow and immediately below $T_{\rm N}$ it starts broadening indicating that the P site is experiencing the static internal field in the ordered state through the hyperfine field between the P nuclei and the ordered Cu$^{2+}$ moments. Below about 4~K, it becomes nearly rectangular with the center of gravity at 1.44~T, which is the resonance field for the $^{31}$P nuclei. It is well documented that for a polycrystalline sample, in a commensurate ordered state, the direction of the internal field is randomly distributed with respect to the applied field and one gets a rectangular spectral shape~\cite{Yamada1986JPSJ,Kikuchi2000JPSJ}. Indeed, such rectangular spectra have been observed in several compounds in the commensurate AFM ordered state. On the other hand, for an incommensurate ordered state, the NMR line shape is expected to be nearly triangular~\cite{Kontani1975JPSJ}. In an attempt to fit our spectrum at 2.8~K, we followed the procedure adopted in Ref.~\onlinecite{Ranjith2016PRB}. The simulated spectra (see the solid line in Fig.~\ref{Fig9}) reproduce the experimental line shape very well, confirming the commensurate nature of the ordering below $T_{\rm N}$.

\begin{figure}
	\includegraphics [scale=1]{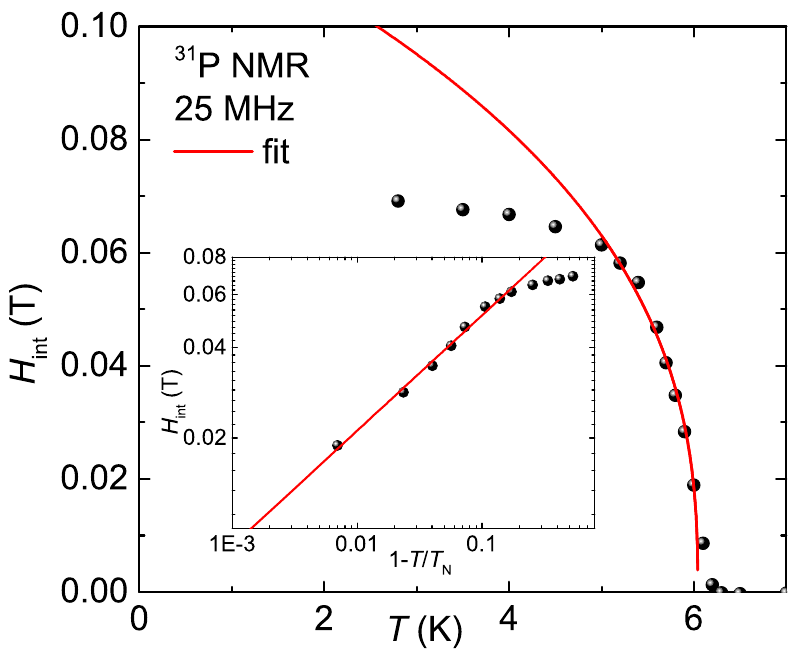}
	\caption{\label{Fig10} (Color online) Temperature dependence of the internal field $H_{\rm int}$ obtained from NMR spectra measured at 25\,MHz in the ordered state. The solid line is the fit by Eq.~\eqref{ms} as described in the text. Inset: $H_{\rm int}$ vs $\tau$.}
\end{figure}
The internal field $H_{\rm int}$, which is proportional to the Cu$^{2+}$ sublattice magnetization, was determined by taking the half-width at half-maximum. The temperature dependence of $H_{\rm int}$ is plotted in Fig.~\ref{Fig10}. In order to extract the critical exponent ($\beta$) of the order parameter, $H_{\rm int}(T)$ was fitted by the power law of the form:
\begin{equation}
	H_{\rm int}(T)=H_{0}\left(1-\frac{T}{T_{\rm N}}\right)^{\beta}.
	\label{ms}
\end{equation}
$H_{\rm int}$ decreases rapidly on approaching $T_{\rm N}$. For an accurate estimation of $\beta$, one needs more data points close to $T_{\rm N}$. We have estimated $\beta$ by fitting the data points in the critical region (close to $T_{\rm N}$) as shown in Fig.~\ref{Fig10}. The value of $\beta \simeq 0.38$ with $H_0 \simeq 0.124$~T and $T_{\rm N} \simeq 6.04$\,K was obtained by fitting the data points in the $T$-range 5\,K to 6\,K, close to $T_{\rm N}$. In order to highlight the fit in the critical region, $H_{\rm int}$ is plotted against the reduced temperature, $\tau = 1-\frac{T}{T_{\rm N}}$ in the inset of Fig.~\ref{Fig10}. The solid line is the fit by $0.124 \times \tau^{0.38}$ where $T_{\rm N}$ is taken to be 6.04~K. At low temperatures, $H_{\rm int}$ develops the tendency toward saturation and saturates faster than expected from the mean-field theory [see the deviation of fits in Fig.~\ref{Fig10} at low-$T$s].

Usually, the critical exponent $\beta$ reflects the universality class of a spin system. The $\beta$ values expected for different spin- and lattice-dimensionalities are listed in Ref.~\onlinecite{Nath2009PRB}. Our experimental value of $\beta \approx 0.38$ is somewhat higher than any of the 3D spin models (Heisenberg, Ising, or XY) suggesting that the magnetic ordering is not driven by simple 3D correlations. This further corroborates our previous assessment based on the $C_{\rm p}(T)$ data.

\subsubsection{Nuclear spin-lattice relaxation rate $1/T_{1}$}
\begin{figure}
	\includegraphics [scale=1]{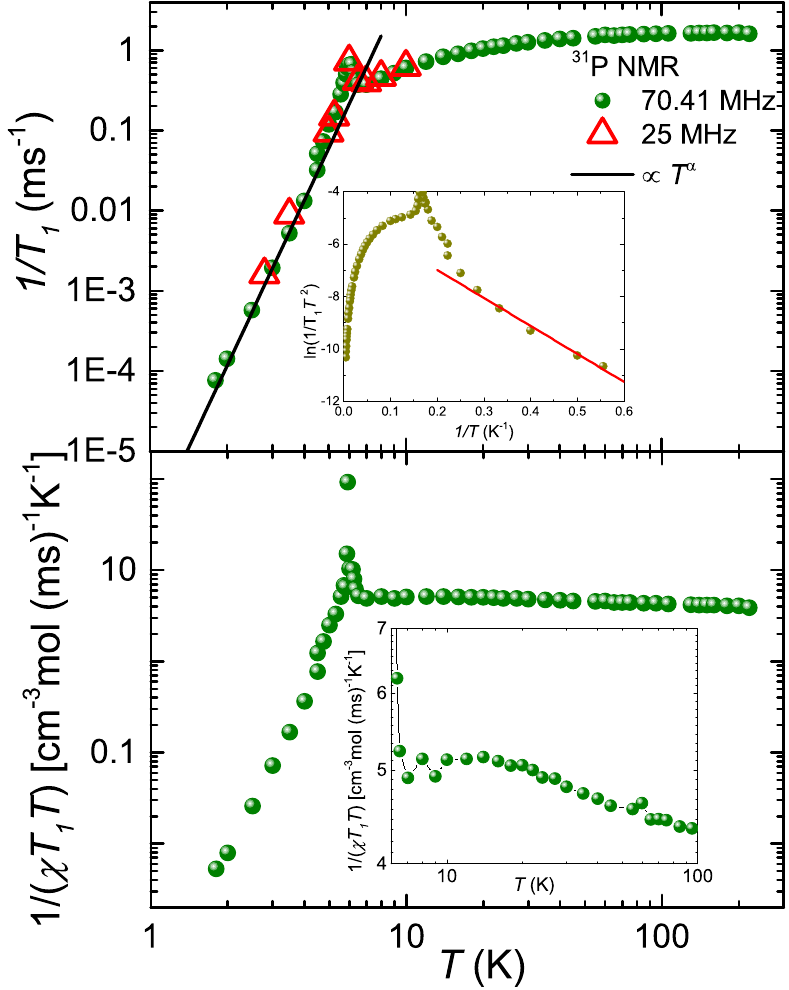}
	\caption{\label{Fig11} (Color online) Upper panel: spin-lattice relaxation rate $1/T_{1}$ vs temperature $T$ measured at 70.41 and 25\,MHz. The solid line represents the fit in the low-temperature region by $1/T_1 \propto T^\alpha$ with $\alpha \simeq 6.8$. Inset: ln($1/T_{1}T^{2}$) vs $1/T$ and the solid line is the linear fit below $T_{\rm N}$. Lower panel: $1/(\chi T_{1}T)$ is plotted as a function of $T$ for 70.41~MHz. Inset: magnified $1/(\chi T_{1}T)$ vs $T$ plot above $T_{\rm N}$.}
\end{figure}
The $^{31}$P nuclear spin-lattice relaxation rate $1/T_{1}$ above $T_{\rm N}$ was measured at the field corresponding to the central peak position. For an $I=1/2$ nucleus, the recovery of the longitudinal magnetization is expected to follow a single-exponential behavior. Indeed, our recovery curves were fitted well by the exponential function
\begin{equation}
1-\frac{M(t)}{M_{0}}=Ae^{-t/T_{1}},
\label{exo}
\end{equation}
where $M(t)$ is the nuclear magnetization at a time $t$ after the saturation pulse and $M_{0}$ is the equilibrium magnetization. The temperature dependence of $1/T_{1}$ extracted from the fit is presented in Fig.~\ref{Fig11}.

The $1/T_{1}$ data measured at two different frequencies (70.41\,MHz and 25\,MHz) almost resemble each other at low temperatures. At high temperatures ($T \gtrsim 30$~K), $1/T_{1}$ is temperature-independent. In the paramagnetic limit $T\gg J/k_{\rm B}$, a temperature-independent $1/T_{1}$ behavior is typically expected due to uncorrelated moments~\cite{Moriya1956}. As the temperature is lowered, $1/T_{1}$ decreases slowly for $T<20$~K and then shows a weak anomaly around $T_{\rm N}\simeq 6.2$~K. Similar decrease has also been reported earlier for some low-dimensional antiferromagnets Pb$_{2}$VO(PO$_{4}$)$_{2}$~\cite{Nath2009PRB}], SrZnVO(PO$_4$)$_2$~\cite{Bossoni2011PRB}, VOMoO$_{4}$~\cite{Carretta2002PRB}, and [Cu(HCO$_{2}$)$_{2}$.4D$_{2}$O], where the decrease of $1/T_{1}$ above $T_{\rm N}$ is explained by the cancellation of the antiferromagnetic spin fluctuations at the probed nuclei~\cite{Carretta2000PRL}. Below $T_{\rm N}$, $1/T_{1}$ again decreases smoothly towards zero.

In the lower panel of Fig.~\ref{Fig11}, $1/(\chi T_{1}T)$ is plotted against temperature. For $T\geq 30$~K, it shows a temperature-independent behavior, and a slow increase was observed just below 30~K where the system begins to show an AFM short-range order. In the inset of Fig.~\ref{Fig11}, the data above $T_{\rm N}$ is magnified in order to highlight the slow increase. The general expression for $1/(T_{1}T)$ in terms of the dynamic susceptibility $\chi_{M}(\mathbf{q},\omega_{0})$ can be written as~\cite{Moriya1963,Mahajan1998PRB}
\begin{equation}
\frac{1}{T_{1}T} = \frac{2\gamma_{N}^{2}k_{B}}{N_{\rm A}^{2}}
\sum\limits_{\vec{q}}\mid A(\mathbf{q})\mid
^{2}\frac{\chi^{''}_{M}(\mathbf{q},\omega_{0})}{\omega_{0}},
\label{t1form}
\end{equation}
where the sum is over wave vectors $\vec{q}$ within the first Brillouin zone, $A(\mathbf{q})$ is the form-factor of the hyperfine interactions as a function of $\mathbf{q}$, and $\chi^{''}_{M}(\mathbf{q},\omega _{0})$ is the imaginary part of the dynamic susceptibility at the nuclear Larmor frequency $\omega _{0}$. For $q=0$ and $\omega_{0}=0$, the real component of $\chi_{M}^{'}(\mathbf{q},\omega _{0})$ corresponds to the uniform static susceptibility $\chi$. Thus the temperature-independent $1/(\chi T_{1}T)$ above 30\,K in Fig.~\ref{Fig11} (bottom panel) demonstrates the dominant contribution of $\chi$ to $1/T_{1}T$. On the other hand, a slight increase in $1/(\chi T_{1}T)$ below 30\,K indicates the growth of AFM correlations with decreasing $T$, which is typically observed in frustrated magnets.

In the magnetically ordered state ($T<T_{\rm N}$), the strong temperature dependence of $1/T_1$ is a clear signature of the relaxation due to scattering of magnons by the nuclear spins~\cite{Belesi2006PRB}. For $T\gg\Delta/k_{\rm B}$, $1/T_1$ follows either a $T^3$ behavior or a $T^5$ behavior due to a two-magnon Raman process or a three-magnon process, respectively, where $\Delta$ is the energy gap in the spin-wave excitation spectrum~\cite{Beeman1968PRB,Nath2014PRB}. At very low temperatures ($T\ll\Delta/k_{\rm B}$), it follows an activated behavior $1/T_1 \propto T^2e^{-\Delta/k_{\rm B}T}$. As shown in the upper panel of Fig.~\ref{Fig11}, $1/T_1$ below $T_{\rm N}$ follows a power law ($\propto T^\alpha$) with an unusually large value $\alpha \simeq 6.8$.
In order to check whether there may exist a magnon gap, we have plotted ln($1/T_{1}T^{2}$) vs $1/T$ in the upper inset of Fig.~\ref{Fig11}. Indeed, it shows a linear regime below $\sim 3.5$~K and a straight line fit below 3.5~K yields the gap value $\Delta/k_{\rm B} \simeq 10.6$~K. This is consistent with the unusually high power-law in the heat capacity, although, as previously mentioned, no clear activated behavior could be seen in $C_{\rm p}(T)$.

\subsection{Microscopic magnetic model}
Exchange couplings were calculated for the experimental crystal structure reported in Ref.~\onlinecite{Kimura2016IC}. First, we mapped valence bands formed by the half-filled Cu orbitals onto a tight-binding model and estimated the hopping parameters $t_i$ (Table~\ref{tab:exchange}), which define AFM superexchange interactions as $J_i^{\rm AFM}=4t_i^2/U_{\rm eff}$, where $U_{\rm eff}$ is the effective on-site Coulomb repulsion in the Cu $3d$ bands. This way, the overall span of the exchange couplings is probed.

\begin{figure}
\includegraphics{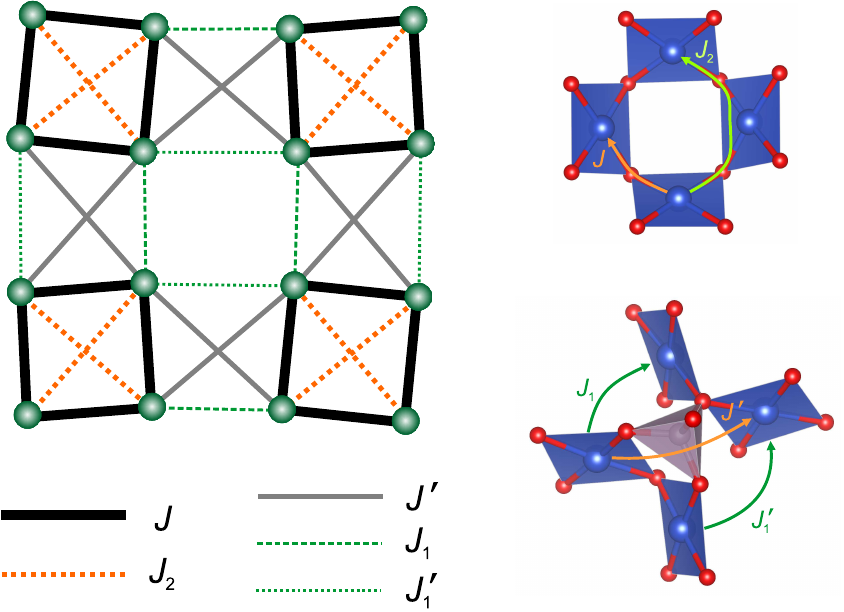}
\caption{\label{fig:lattice}
Spin lattice of Sr(TiO)Cu$_{4}$(PO$_{4}$)$_{4}$ and mechanisms of individual exchange couplings. Note that the \mbox{Cu--O$\ldots$O--Cu} superexchange pathway of $J'$ is more straight than those of $J_1$ and $J_1'$. 
}
\end{figure}
The spin lattice of Sr(TiO)Cu$_{4}$(PO$_{4}$)$_{4}$ comprises square cupola with the leading nearest-neighbor coupling $J$. A weaker and frustrating second-neighbor coupling $J_2$ may exist too. Several superexchange pathways are present between the cupola. Among them, the second-neighbor ("diagonal") $J'$ is clearly more efficient than the nearest-neighbor $J_1$ and $J_1'$ (Fig.~\ref{fig:lattice}). The leading interplane coupling runs perpendicular to the $ab$ plane via $t_{\perp}\simeq 5$\,meV that corresponds to a minute $J_{\perp}^{\rm AFM}\simeq 0.3$\,K assuming $U_{\rm eff}=4$\,eV~\cite{nath2013}.

\begin{table}
\caption{\label{tab:exchange}
Calculated exchange couplings in Sr(TiO)Cu$_{4}$(PO$_{4}$)$_{4}$: the interatomic distances $d_{\rm Cu-Cu}$ (in\,\r A), hopping parameters $t_i$ (in\,meV), and total exchange couplings $J_i$ (in\,K) obtained from DFT+$U$ with $U_d=9.5$\,eV and $J_d=1$\,eV.
}
\begin{ruledtabular}
\begin{tabular}{ccrc}
    & $d_{\rm Cu-Cu}$ & $t_i$ & $J_i$ \\
$J$ & 3.147 & 118 & 27.9 \\
$J_1$ & 3.619 & 23 & 1.0 \\
$J_1'$ & 4.101 & $-9$ & 0.0 \\
$J_2$ & 4.450 & 14 & 5.8 \\
$J'$ & 4.973 & $-48$ & 7.0 \\
$J_2'$ & 5.097 & 10 & 2.2 \\
\end{tabular}
\end{ruledtabular}
\end{table}

DFT+$U$ calculations give access to absolute values of the exchange couplings, including both FM and AFM contributions (Table~\ref{tab:exchange}). They confirm the leading coupling $J$, the presence of the frustrating second-neighbor coupling $J_2$ within the cupolas and the non-frustrated coupling $J'$ between them. Therefore, the minimum magnetic model comprises isolated square cupolas with only the nearest-neighbor coupling $J$. As a better approximation, the couplings $J_2$ and $J'$ should be included. 

\begin{figure}
\includegraphics{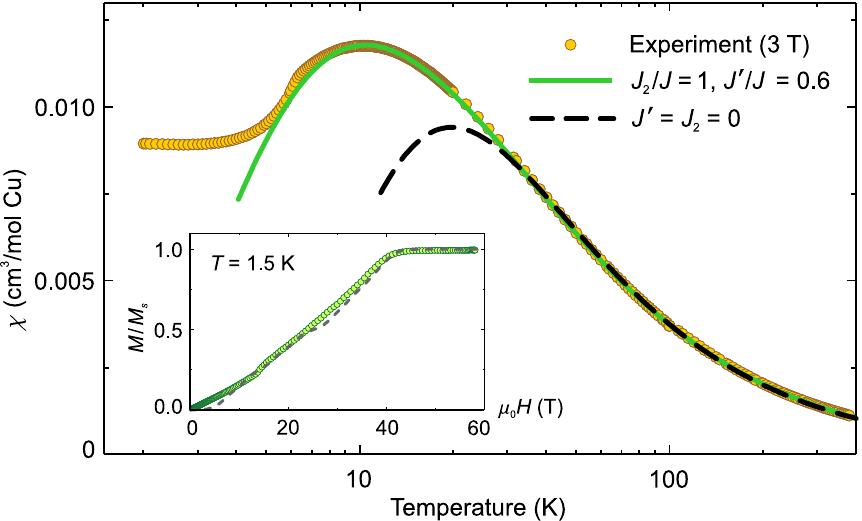}
\caption{\label{fig:chi-fit}
Magnetic susceptibility of Sr(TiO)Cu$_{4}$(PO$_{4}$)$_{4}$ (circles) and the fits within the models of isolated square cupolas ($J=25.3$\,K, $J_2=J'=0$, $g=2.14$, dashed line) and coupled frustrated cupolas ($J=16.3$\,K, $J_2/J=1.0,J'/J=0.6$, $g=2.17$, solid line). The inset shows experimental magnetization curve (circles) and the simulation using the latter model at $T/J=0.1$ (short-dashed line). The bends in the simulated curve are due to finite-size effects.}
\end{figure}

The model of isolated cupolas with only nearest-neighbor couplings describes the susceptibility down to 30\,K but clearly fails to reproduce the position and height of the maximum (Fig.~\ref{fig:chi-fit}, dashed line). Including $J'$ improves the fit down to about 20\,K. However, the main improvement is achieved by including $J_2$ that, even in the absence of $J'$, describes the data down to 14\,K. By combining $J'$ and $J_2$, we get the best fit using $J=16.3$\,K, $J_2/J=1.0$, and $J'/J=0.6$ down to 8.5\,K. Both position and height of the maximum are well reproduced (Fig.~\ref{fig:chi-fit}, solid line). The fitted $g$-value of 2.17 seems realistic for Cu$^{2+}$ and consistent with $g=2.18$ from the Curie-Weiss fit. 

Same exchange parameters were used to simulate the magnetization process of Sr(TiO)Cu$_{4}$(PO$_{4}$)$_{4}$ and lead to a good agreement with the experimental saturation field (Fig.~\ref{fig:chi-fit}, inset). However, finite-size effects are significant, and not all features of the experimental curve are reproduced, because the transition around 15\,T is anisotropic in nature~\cite{Kato2017PRL} and can not be captured on the level of Heisenberg spin Hamiltonian. 

\section{Discussion and Conclusions}
Similar to BTCPO, the spin lattice of STCPO comprises square cupolas. However, the nearest-neighbor coupling within the cupolas is by far insufficient to describe magnetic susceptibility of this compound. Our fits show that the second-neighbor coupling $J_2$ and the coupling $J'$ between the cupolas are both sizable and integral to the magnetic model. 

The moderate size of $J$, 27.9\,K in DFT and 16.3\,K in the experiment, is controlled by the Cu--O--Cu angle of $107.4^{\circ}$~\cite{Kimura2016IC} that lies on the border between FM and AFM superexchange. In this range of the bridging angles, subtle structural details, as well as side groups, have strong influence on the absolute value and even on the sign of the coupling~\cite{nath2013,lebernegg2017,badrtdinov2018}. In BTCPO, a somewhat larger nearest-neighbor coupling of $J=35$\,K was reported~\cite{Kimura2016NComm,Kato2017PRL}. Indeed, the Ba compound systematically shows higher values of $T_N\simeq 9.5$\,K~\cite{Kimura2016NComm} (vs. 6.2\,K), $T_{\chi}^{\max}\simeq 15$\,K~\cite{Kimura2016NComm} (vs. 10\,K), and $H_s\simeq 60$\,T~\cite{Kato2017PRL} (vs. 40\,T). Here, the values in brackets are for the Sr compound reported in this work.

The second-neighbor coupling $J_2$ is more unusual, because its experimental value of about 16\,K is much larger than 5\,K found in DFT. One plausible explanation for this discrepancy would be subtle structural changes upon cooling, as room-temperature crystal structure was used in DFT calculations for the lack of any crystallographic information at lower temperatures. The mechanism of $J_2$ should involve Cu--O$\ldots$O--Cu superexchange mediated by the relatively short O$\ldots$O distance of 2.76\,\r A along the edge of the CuO$_4$ plaquette. A similar mechanism should be relevant to the couplings $J'$, $J_1$, and $J_1'$, but the shorter O$\ldots$O contact of $2.48-2.51$\,\r A is along the edge of the PO$_4$ tetrahedron in this case. The size of the coupling is controlled by deviations from linearity quantified by the deviation of the Cu--O--O angles from $180^{\circ}$~\cite{tsirlin2010,nath2014,berdonosov2013}. The strongest coupling $J'$ is found for the least curved Cu--O$\ldots$O--Cu superexchange pathway (Fig.~\ref{fig:lattice}).

Magnetic response of STCPO reveals several peculiarities. First, the magnetic ordering transition at $T_N\simeq 6.2$\,K is observed well below the susceptibility maximum at $T_{\chi}^{\max}\simeq 10$\,K, and the ratio $\theta_{\rm CW}/T_N=3.0$ indicates a sizable reduction in $T_N$ due to the low-dimensionality and frustration. Second, the long-range-ordered state appears to be commensurate, but both specific heat and $1/T_1$ decrease rapidly at low temperatures. This may indicate gapped nature of magnetic excitations.

A somewhat different coupling regime was reported for BTCPO in previous studies. The authors of Refs.~\onlinecite{Kimura2016NComm,Kato2017PRL} argue that $J'$ is about 50\% of $J$, whereas $J_2$ is only a minor coupling ($J_2/J=\frac16$). We note, however, that these values were not tested against the experimental magnetic susceptibility, and only a classical analysis of the spin model was performed~\cite{Kato2017PRL}. Gapped magnetic excitations may be related to the non-neglibigle anisotropy postulated for the Ba compound. On the other hand, the coupling $J_2$ can play a role in the stabilization of the non-coplanar state, an effect hitherto ascribed to the Dzyaloshinsky-Moriya anisotropy~\cite{Kato2017PRL}. 

In summary, we explored the low-temperature behavior of Sr(TiO)Cu$_{4}$(PO$_{4}$)$_{4}$, a quantum magnet comprising the square cupolas of Cu$^{2+}$ ions. Our results indicate sizable couplings between individual cupola units, as well as strong frustration within the cupolas. A spin gap of about 10\,K may form in the ordered state, resulting in the unusually high power-law exponent of the specific heat and the activated behavior of the spin-lattice relaxation rate at low temperatures.

{\it Note Added}: after the submission of our work, Kimura {\it et al.}~\cite{Kimura2018PRB} reported the magnetic structure of STCPO from neutron diffraction experiments. Their commensurate magnetic structure is consistent with our NMR data. Additionally, STCPO shows a pronounced dielectric anomaly at $T_{\rm N}$ but no ferroelectricity was observed.

\acknowledgments
SSI and RN would like to acknowledge BRNS, India for financial support bearing sanction No.37(3)/14/26/2017-BRNS. AT acknowledges financial support by the Federal Ministry for Education and Research via the Sofja Kovalevskaya Award of Alexander von Humboldt Foundation.


%

\end{document}